\documentclass[reprint,amsmath,amssymb,aps]{revtex4-2}

\newcommand{\RomanNumeralCaps}[1]
    {\MakeUppercase{\romannumeral #1}}

\usepackage{graphicx}
\usepackage{hyperref}
\usepackage{siunitx}
\usepackage{dcolumn}
\usepackage{bm}
\usepackage{color}
\usepackage{xcolor}
\usepackage{needspace}
\DeclareUnicodeCharacter{2212}{-}

\begin{document}


\title{Competing Orders Driven by Wigner Crystal Phase in Rhombohedral Graphene}

\author{Zekang Zhou$^{1,^*}$}
\author{Kilian Kr\"otzsch$^{1,^*}$}
\author{Rapha\"el Ayache$^{2}$}
\author{Yonggen Li$^{1}$}
\author{Sandeep Joy$^{3,4, 5}$}
\author{Kenji Watanabe$^{6}$}
\author{Takashi Taniguchi$^{7}$}
\author{Moty Heiblum$^{8}$}
\author{Preden Roulleau$^{2}$}
\author{Mitali Banerjee$^{1,9, \dagger}$}

\affiliation{$^{1}$Institute of Physics, Ecole Polytechnique Fédérale de Lausanne, Switzerland}

\affiliation{$^{2}$SPEC, CEA, CNRS, Université Paris-Saclay, CEA Saclay, 91191 Gif sur Yvette Cedex
France}

\affiliation{$^{3}$Department of Physics, Florida State University, Tallahassee, FL 32306}

\affiliation{$^{4}$National High Magnetic Field Laboratory, Tallahassee, FL 32310}

\affiliation{$^{5}$ FSU Quantum Initiative, Florida State University, Tallahassee, FL 32306}

\affiliation{$^{6}$Research Center for Functional Materials, National Institute for Materials Science, Japan}

\affiliation{$^{7}$International Center for Materials Nanoarchitectonics, National Institute for Materials Science, Japan}

\affiliation{$^{8}$Braun Center for Submicron Research, Department of Condensed Matter Physics, Rehovot, Israel}

\affiliation{$^{9}$Center for Quantum Science and Engineering, Ecole Polytechnique Fédérale de Lausanne, Switzerland}
 
\affiliation{$^{\dagger}$ Email: mitali.banerjee@epfl.ch}

\affiliation{$^*$ These authors contributed equally}

\begin{abstract}

Rhombohedral graphene systems provide a unique platform where strong electronic interactions and nontrivial band topology coexist at low carrier densities and high displacement fields, giving rise to a rich landscape of emergent electronic phases. Here, we report that the highly insulating state on the low-density side of chiral superconductivity in rhombohedral pentalayer graphene (R5G) corresponds to a Wigner crystal (WC) phase. In addition, a hole-doped metallic Wigner crystal (h-mWC) phase emerges near the WC boundary. Under an out-of-plane magnetic field, the system hosts competing magnetic-field-stabilized superconductivity (fSC) and unconventional reentrant quantum Hall (RIQH) states. These emergent phases are closely connected to the underlying WC and mWC states and evolve continuously across phase boundaries. Our results establish that WC phase plays an important role in the phase diagram of rhombohedral multilayer graphene and highlight its connection to a rich landscape of emergent phases.

\end{abstract}

\maketitle

\begin{figure*}
  \centering
  \includegraphics[width= 1\textwidth]{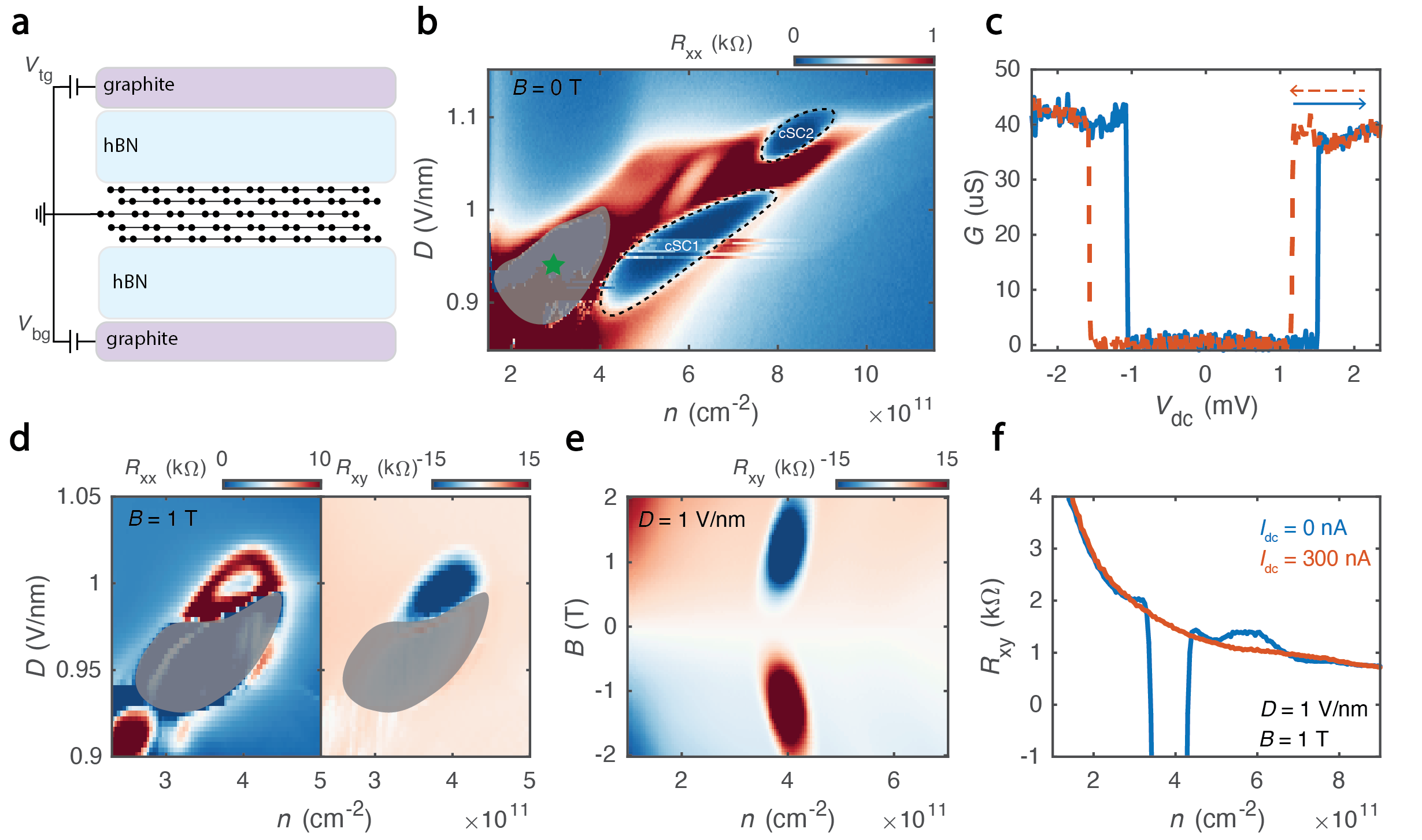} \caption{\textbf{Wigner and metallic Wigner crystals in R5G.} \textbf{a,} Schematic of the dual-graphite-gated device geometry that enables independent tuning of the carrier density and the displacement field. \textbf{b,} $n$--$D$ map of $R_{xx}$ at $B = 0\,\mathrm{T}$ ($T = 20\,\mathrm{mK}$). Two chiral superconducting states (cSC1 and cSC2) are visible in this region. At low carrier densities, we observe a region of extremely high resistance, highlighted by the light-gray shaded area, where current injection into the system is suppressed. \textbf{c,} Conductance as a function of DC bias voltage at the point marked by the green star in \textbf{b}. The blue and orange curves represent forward and backward sweeps, respectively. Shape switching and hysteresis are observed ($T = 20\,\mathrm{mK}$). \textbf{d,} $n$--$D$ maps of $R_{xx}$ and $R_{xy}$ at $B = 1\,\mathrm{T}$, obtained by symmetrizing and antisymmetrizing the data measured at $B = \pm 1\,\mathrm{T}$ ($T = 20\,\mathrm{mK}$). \textbf{e,} Landau fan diagram of $R_{xy}$ (antisymmetrized) at $D = 1\,\mathrm{V/nm}$ ($T = 240\,\mathrm{mK}$). \textbf{f,} Hall resistance as a function of carrier density measured at $I_{\mathrm{dc}} = 0~\mathrm{nA}$ and $300~\mathrm{nA}$ ($T = 240\,\mathrm{mK}$).}
  \label{fig1}
\end{figure*}

In a two-dimensional electron gas, lowering the carrier density increases the ratio of Coulomb interaction energy to kinetic energy, pushing the system from a Fermi liquid toward a strongly correlated state. In the strong-coupling limit, the system spontaneously breaks translational symmetry and localizes electrons to form a Wigner crystal~\cite{wigner1934interaction}, a paradigmatic interaction-driven electronic solid that has been extensively pursued experimentally in different systems~\cite{grimes1979evidence, andrei1988observation, yoon1999wigner, chen2019competing, falson2022competing, smolenski2021signatures,zhou2021bilayer,tsui2024direct, xiang2025imaging, ge2025visualizing}. When the quantum geometry of the underlying electronic band is taken into account, the Wigner crystal exhibits additional emergent properties beyond the conventional interaction-driven picture. First, Berry curvature can endow the electronic orbitals in a Wigner crystal with chirality, resulting in a finite orbital angular momentum associated with localized electrons~\cite{joy2025chiral, soejima2025lambda}. The geometric effects can also manifest at the level of the many-body crystalline state, leading to nontrivial topological properties of the Wigner crystal. In particular, Berry curvature can enable a topological Wigner crystal that breaks translational symmetry while retaining a nonzero Chern number~\cite{tevsanovic1989hall, zhou2024fractional, soejima2024anomalous, dong2024anomalous, dong2024stability,zeng2024sublattice, tan2024parent, kwan2025moire, valenti2025quantum, desrochers2026electronic}. Magnetic field provides an additional control parameter through orbital quantization and exchange interactions, reshaping the competition among phases and enabling access to additional phases. The physics becomes even richer as the Wigner crystal approaches melting upon tuning the electron density. Near the transition, competing instabilities may emerge, such as the microemulsion regime with coexisting Wigner crystal and liquid domains~\cite{spivak2004phases,spivak2010transport,jamei2005universal, joy2023upper}, and the metallic Wigner crystal where crystalline order coexists with self-doping induced itinerant carriers~\cite{andreev1969quantum, fisher1979defects, cockayne1991energetics, pankov2008self, kim2024dynamical, dong2026crystals, feng2026self}. However, experimental platforms for systematically exploring the interplay among strong correlations, nontrivial band topology, magnetic-field effects remain limited.

In this context, rhombohedral multilayer graphene (RG) has emerged as a particularly promising platform, where displacement-field-tunable flat bands enable access to strongly correlated electron systems with nontrivial band topology. Notably, experimental studies have already reported the observation of the fractional quantum anomalous Hall effect (FQAHE)~\cite{lu2024fractional,xie2025tunable} as well as chiral superconductivity (cSC)~\cite{han2025signatures} in RG systems with and without the introduction of a moiré superlattice, respectively. Irrespective of the alignment with hexagonal boron nitride (hBN), these systems exhibit a highly insulating state at electron densities below those of the FQAHE and cSC phases~\cite{lu2024fractional,xie2025tunable,han2025signatures,lu2025extended}. Understanding this highly insulating state is therefore crucial for understanding the nature of the exotic phases in RG systems. In this work, we show that the highly insulating state in rhombohedral pentalayer graphene (R5G) corresponds to a Wigner crystal phase, which underlies a rich landscape of emergent phenomena, including superconductivity and reentrant quantum Hall states.

\Needspace{6\baselineskip}
\section{Wigner crystal and metallic Wigner crystal in R5G}

\begin{figure*}
  \centering
  \includegraphics[width= 0.85\textwidth]{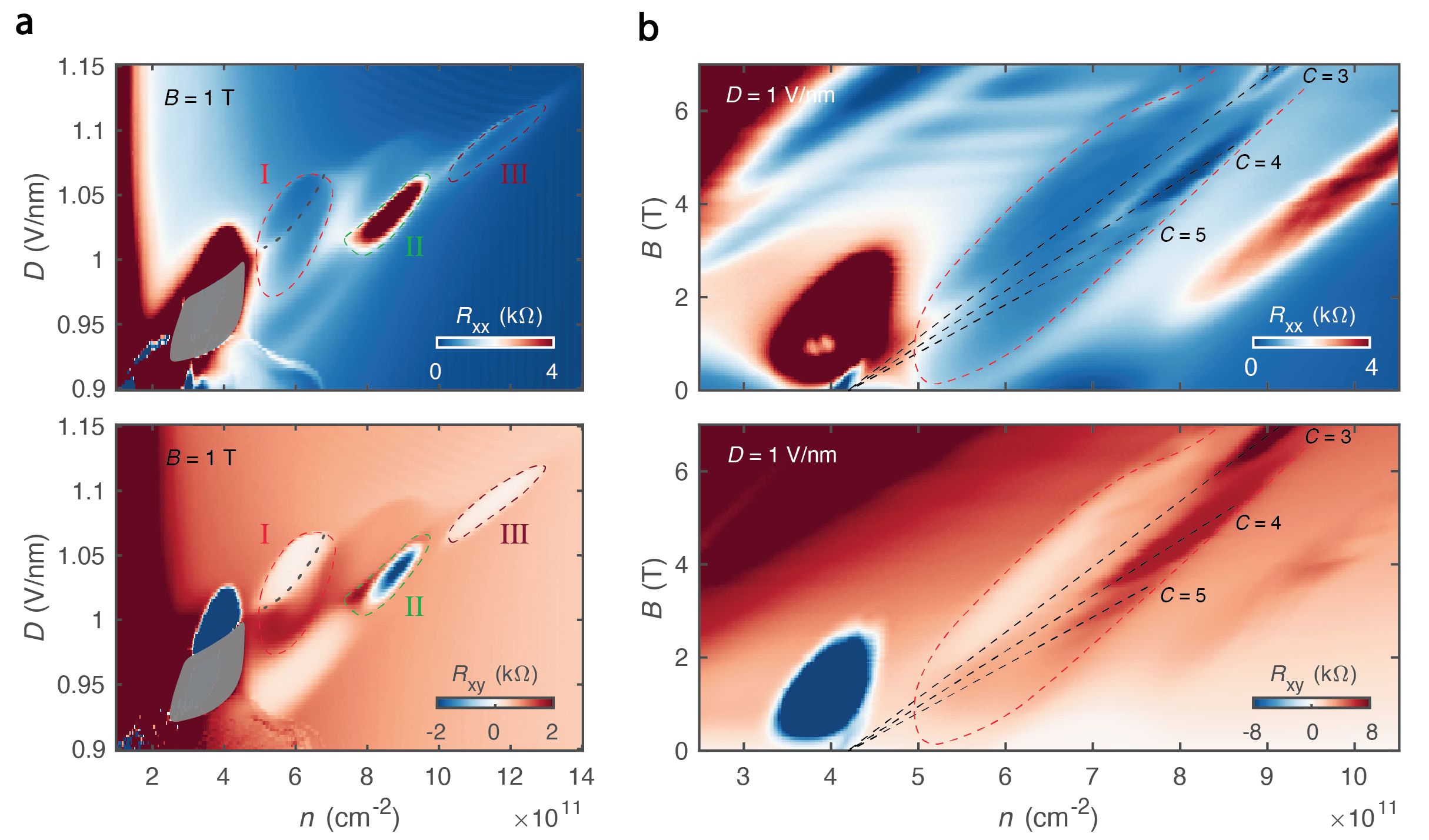} \caption{\textbf{Bifurcation of region \RomanNumeralCaps{1}.} \textbf{a,} $n$--$D$ maps of $R_{xx}$ and $R_{xy}$ at $B = 1\,\mathrm{T}$, obtained by symmetrizing and antisymmetrizing the data measured $B = \pm 1\,\mathrm{T}$. Apart from the regions in the phase diagram that host cSC1 and cSC2, we identify three distinct regions, \RomanNumeralCaps{1}, \RomanNumeralCaps{2}, and \RomanNumeralCaps{3}, in the remaining portions of the $n$--$D$ phase space. \RomanNumeralCaps{1} lies above cSC1, \RomanNumeralCaps{2} lies below cSC2, and \RomanNumeralCaps{3} resides at the highest electron density. We will focus on region \RomanNumeralCaps{1}, enclosed by the red dashed line, which is clearly divided into two parts. The black dashed line within region \RomanNumeralCaps{1} marks the boundary between these two distinct parts. \textbf{b,} Landau fan diagram of $R_{xx}$ and $R_{xy}$ at $D = 1\,\mathrm{V/nm}$. The red dashed line marks the boundary of region \RomanNumeralCaps{1} in the fan diagram. In addition to conventional Landau levels, RIQH states are observed at the higher electron-doping side in region \RomanNumeralCaps{1}, as indicated by the black dashed lines. These RIQH states can be traced back to the boundary of the WC state in the fan diagram. ($T = 10\,\mathrm{mK}$)}
  \label{fig2}
\end{figure*} 

We fabricated dual-graphite-gated pentalayer rhombohedral graphene (R5G) devices encapsulated in hBN, as shown in Fig.~\ref{fig1}a. This geometry enables independent and continuous control over the carrier density $n$ and the displacement field $D$. This work investigates two R5G devices, R5G\_1 and R5G\_2. The main text focuses on the results from R5G\_1, while the results from R5G\_2 are presented in the Extended Data. Fig.~\ref{fig1}b shows the $n$--$D$ map of $R_{xx}$ in the high-displacement-field region. Two zero-resistance regions enclosed by black dashed lines are visible and were identified in Ref.~\cite{han2025signatures} as chiral superconducting phases (cSC1 and cSC2). On the low carrier density side of the cSC1, an insulating region emerges, where the resistance becomes unmeasurable due to the inability to inject current into the highly insulating state (marked by the light-gray shaded area). Similar behavior has been reported in other RG systems~\cite{han2025signatures,lu2024fractional}, where it appears adjacent to cSC and the FQAHE states. 

Such robust insulating behavior in a non-moiré crystalline system with flat bands suggests the possible formation of a Wigner crystal phase. To further probe the true nature of this insulating state, we perform DC bias voltage-dependence measurements at the point marked by the green star in Fig.~\ref{fig1}b. They are shown in Fig.~\ref{fig1}c. The conductance is measured while sweeping the DC voltage bias. At low DC bias voltage, the conductance is nearly zero, consistent with a highly insulating state. Beyond a threshold bias, the conductance exhibits a sharp increase. Crucially, the forward and backward bias sweeps exhibit a pronounced hysteretic loop in conductance. Such threshold behavior and hysteresis are characteristic transport signatures of a Wigner crystal and reflect collective pinning by disorder and subsequent depinning under a critical driving electric field~\cite{littlewood1987pinning, druner1988the, williams1991conduction, willett1989current, csathy2007astability, munyan2024evidence}. In hBN-encapsulated devices, the primary source of disorder is likely Coulomb disorder from the substrate. Fig.~\ref{fig1}d shows the $n$--$D$ maps of $R_{xx}$ and $R_{xy}$ at $B = 1\,\mathrm{T}$, where $R_{xx}$ is symmetrized and $R_{xy}$ is antisymmetrized with respect to the magnetic field. The WC phase remains clearly visible, as marked by the light-gray shaded area. Near the upper boundary of the WC phase, a distinct phase emerges, characterized by a dip in $R_{xx}$ and a sign reversal in $R_{xy}$, consistent with recent reports of a mWC phase in RG systems~\cite{han2026evidence}. In the mWC state, the self-doping induced itinerant carriers coexist with the pinned background electron lattice, and the itinerant carriers may be either electron-like or hole-like, depending on the details of the system~\cite{andreev1969quantum,kim2024dynamical, dong2026crystals,feng2026self}. The phase observed here is a hole-doped electron WC, as evidenced by its negative Hall resistance. Fig.~\ref{fig1}e presents the Landau fan diagram of $R_{xy}$ across the mWC region. A clear sign reversal of the Hall signal is observed, and the carrier density at which the sign change occurs shifts with the applied magnetic field. Notably, the sign reversal of $R_{xy}$ is observed only in the presence of a finite magnetic field, suggesting that the mWC phase is stabilized by the magnetic field and then disappears at high magnetic field. The nature of the mWC state in a magnetic field warrants further investigation. We examined the Hall response under a high DC bias current. Fig.~\ref{fig1}f shows the Hall resistance measured at different bias currents. The blue curve ($I_{\mathrm{dc}} = 0$) exhibits a clear sign change in $R_{xy}$, consistent with the mWC phase and its anomalous carrier response. In contrast, the red curve ($I_{\mathrm{dc}} = 300~\mathrm{nA}$) displays a conventional electron-like Hall response (Fig.~\ref{E1}a shows the nonlinear behavior of the mWC phase as a function of carrier density). Elevated temperatures also drive $R_{xy}$ back to an electron-type sign due to melting of the underlying WC phase (Fig.~\ref{E1}b). Similar behavior has been observed in Ref.~\cite{han2026evidence}. 

\begin{figure*}
  \centering
  \includegraphics[width= 0.95\textwidth]{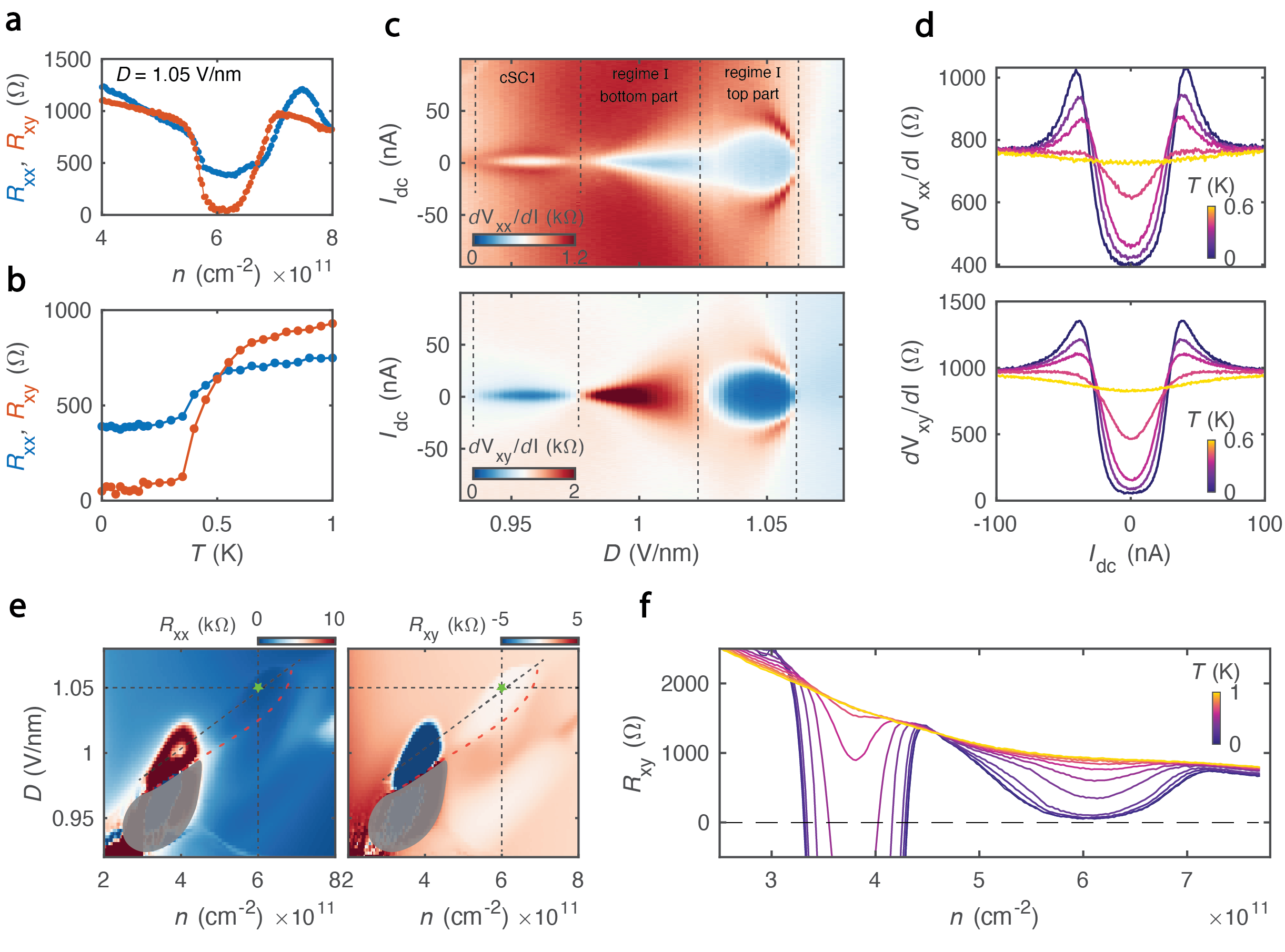} \caption{\textbf{Magnetic-field-stabilized superconductivity near the metallic Wigner crystal.} \textbf{a,} $R_{xx}$ (symmetrized) and $R_{xy}$ (antisymmetrized) as a function of carrier density $n$ at $B = 1\,\mathrm{T}$ and $D = 1.05\,\mathrm{V/nm}$ (horizontal dashed line in \textbf{e}). Both $R_{xx}$ and $R_{xy}$ show a sudden drop, with $R_{xy}$ approaching zero. \textbf{b,} Temperature dependence of $R_{xx}$ and $R_{xy}$ at the optimal doping point in \textbf{a}, corresponding to a carrier density of $n = 0.628 \times 10^{12}\,\mathrm{cm^{-2}}$. \textbf{c}, $dV_{xx}/dI$ and $dV_{xy}/dI$ as a function of DC bias current $I_{dc}$ and displacement field at a fixed carrier density of $n = 0.6 \times 10^{12}$ at $B = 1\,\mathrm{T}$(vertical dashed line in \textbf{e}). \textbf{d}, Temperature dependence of $dV_{xx}/dI$ (symmetrized) and $dV_{xy}/dI$ (antisymmetrized) at $D = 1.05\,\mathrm{V/nm}$ and $n = 0.6 \times 10^{12}$ and $B = 1\,\mathrm{T}$(green star point in \textbf{e}). \textbf{e,} $n$--$D$ maps of $R_{xx}$ and $R_{xy}$ at $B = 1\,\mathrm{T}$. The red dashed line is a guide to the eye marking the boundary between the mWC and WC, as well as the boundary between the two parts in region \RomanNumeralCaps{1}. \textbf{f,} Temperature dependence of $R_{xy}$ along the diagonal black dashed line in \textbf{e}, across the mWC and the magnetic-field-stabilized superconductivity.}
  \label{fig3}
\end{figure*}

\section{Bifurcation of region \RomanNumeralCaps{1}}

Having identified the WC and mWC phases, we next investigate the surrounding regions in the $n$--$D$ phase diagram under an out-of-plane magnetic field. Fig.~\ref{fig2}a shows the $n$--$D$ maps of $R_{xx}$ and $R_{xy}$ over a wide parameter range at $B = 1\,\mathrm{T}$. Both the cSC1 and cSC2 phases are substantially suppressed by the applied magnetic field. Apart from the cSC regions, we identify three distinct regions that do not exhibit conventional Landau level features, labeled \RomanNumeralCaps{1}, \RomanNumeralCaps{2}, and \RomanNumeralCaps{3}. We find that region \RomanNumeralCaps{1}, whose boundary is marked by the red dashed line, is bounded by the lower density WC/mWC phase on the left, the higher density cSC2 phase on the right, and the lower displacement field cSC1 phase below (both cSC states are labeled in Fig.~\ref{fig1}b). Understanding the physics in region \RomanNumeralCaps{1} is therefore essential for clarifying both the origin of the surrounding exotic phases and their interconnections. Within region \RomanNumeralCaps{1}, we identify two distinct parts, with the boundary between them marked by the black dashed line in Fig.~\ref{fig2}a. The contrasting behavior of $R_{xy}$ between these two parts, despite similarly low $R_{xx}$, points to distinct underlying electronic phases.

Fig.~\ref{fig2}b shows the Landau fan diagrams of $R_{xx}$ and $R_{xy}$ at $D = 1\,\mathrm{V/nm}$. At both low and high carrier densities, the system exhibits conventional Landau level features, corresponding to an annular Fermi surface at low carrier density and a simply-connected Fermi surface at high carrier density~\cite{han2025signatures, qin2025extreme} (The high density Landau level features are visible in Fig.~\ref{fig2}a). Region \RomanNumeralCaps{1}, whose boundary is again marked by the red dashed line in Fig.~\ref{fig2}b, shows an unconventional dependence on the magnetic field. We first note that the boundaries of region \RomanNumeralCaps{1} shift to higher carrier density when the magnetic field is increased, and that the upper and lower parts of region \RomanNumeralCaps{1} in the $n$--$D$ correspond to the lower and higher electron-doped sides inside of region \RomanNumeralCaps{1} in the Landau fan diagrams, respectively. Second, confined to the lower part of region \RomanNumeralCaps{1}, we observe RIQH states, marked by the black dashed lines in Fig.~\ref{fig2}b. Similar to conventional quantum Hall states, these RIQH states also exhibit quantized Hall resistance $R_{xy}$ and vanishing $R_{xx}$. However, there are some important differences between the two. In particular, the RIQH states do not originate from the charge neutrality point; instead, they emerge from the WC phase boundary. Moreover, these RIQH states deviate from the St\v{r}eda formula in the fan diagram. The nature of these RIQH states will be discussed in a later section.

\section{superconductivity near the metallic Wigner crystal state}

We now aim to elucidate the nature of the state in the upper part of region \RomanNumeralCaps{1}. Fig.~\ref{fig3}a presents line cuts of $R_{xx}$ and $R_{xy}$ measured at $B = 1\,\mathrm{T}$ and $D = 1.05\,\mathrm{V/nm}$. Both resistances exhibit a large drop, with $R_{xy}$ approaching nearly zero. Consistent behavior is observed in R5G\_2, where $R_{xy}$ similarly decreases from its normal Hall resistance to nearly zero (See Fig.~\ref{E2}), which demonstrates the reproducibility of this feature across devices. Fig.~\ref{fig3}b shows the temperature dependence of $R_{xx}$ and $R_{xy}$ at $B = 1\,\mathrm{T}$. Both resistances saturate at low temperatures and increase rapidly above a characteristic temperature. Fig.~\ref{fig3}c displays $dV_{xx}/dI$ and $dV_{xy}/dI$ as functions of DC bias current and displacement field at a fixed carrier density of $n = 0.6 \times 10^{12}\,\mathrm{cm}^{-2}$. The data can be divided into four distinct regions. The weak nonlinear behavior at low displacement field corresponds to the cSC1 state, which is not completely suppressed by the magnetic field. As the displacement field is increased, the system enters the lower part of region \RomanNumeralCaps{1}, where nonlinear transport persists but no clear characteristics of superconductivity are present. Upon further increasing the displacement field, the system transitions into the upper part of region \RomanNumeralCaps{1}, where both $R_{xx}$ and $R_{xy}$ exhibit low resistance together with pronounced nonlinear behavior and a clear critical current feature. Finally, the system enters a normal state region at high displacement fields, accompanied by the disappearance of nonlinear behavior. Fig.~\ref{fig3}d shows the temperature dependence of the nonlinear transport behavior as a function of DC bias current in the upper part of region \RomanNumeralCaps{1}. The nonlinear response is gradually suppressed with increasing temperature and eventually evolves into linear ohmic behavior.

The sudden drop in resistance as a function of carrier density, the saturation of $R_{xx}$ and the vanishing $R_{xy}$ at low temperature, and the temperature dependence of the nonlinear behavior suggest that the upper part of region\,\RomanNumeralCaps{1} hosts a superconducting state, although a perfectly vanishing $R_{xx}$ is not observed. We note that previous studies of RG systems have reported that some superconducting states do not show perfectly vanishing resistance~\cite{zhou2021superconductivity, yang2025impact, seo2025family, xie2026magnetic}. Interestingly, this combination of a finite $R_{xx}$ and a vanishing $R_{xy}$ is reminiscent of an anomalous metallic phase that has been extensively discussed in disordered superconductivity~\cite{kapitulnik2019colloquium}, and that has recently gained attention in RG systems~\cite{yankowitz2026superconductivity}. The vanishing $R_{xy}$ originates from an emergent particle-hole symmetry inherited from the superconducting state~\cite{breznay2017particle}, a mechanism that may also be at play in the R5G system. Future studies, including shot noise measurements to detect the quasi-particles' charge and other phase-sensitive measurements, will be crucial to elucidate the true origin of this non-zero $R_{xx}$. A recent study on R6G has also reported magnetic-field-stabilized superconductivity that coexists with RIQH states in a similar region of the $n$--$D$ phase diagram~\cite{nguyen2025hierarchy}, suggesting that such observations might be a generic phenomenon across RG systems with different layer numbers. We find that region \RomanNumeralCaps{3} also exhibits a sudden drop in resistance alongside strongly nonlinear $V\text{--}I$ characteristics (Figs.~\ref{E1} and \ref{E4}).

As mentioned above and shown in Fig.~\ref{fig2}, region \RomanNumeralCaps{1} can be divided into two distinct parts. We have demonstrated that the upper part hosts an fSC state, whereas the lower part can support RIQH states, with a clear boundary separating the two phases. In Fig.~\ref{fig1}, we also identified another phase boundary, which separates the mWC phase and the WC phase. Interestingly, these two boundaries are connected to each other, as indicated by the red dashed line in Fig.~\ref{fig3}e. Fig.~\ref{fig3}f shows the temperature dependence of $R_{xy}$ along the diagonal dashed line in Fig.~\ref{fig3}e, crossing the mWC phase and the fSC phase. The less electron-doped side exhibits a negative Hall resistance, corresponding to the mWC state, while the more electron-doped side shows a nearly zero Hall resistance, characteristic of the superconducting state. The two states show similar temperature dependence. Although we cannot yet comment on the detailed relationship between the two states, these observations suggest that they may share a common underlying nature. 

\begin{figure*}
  \centering
 \includegraphics[width= 1\textwidth]{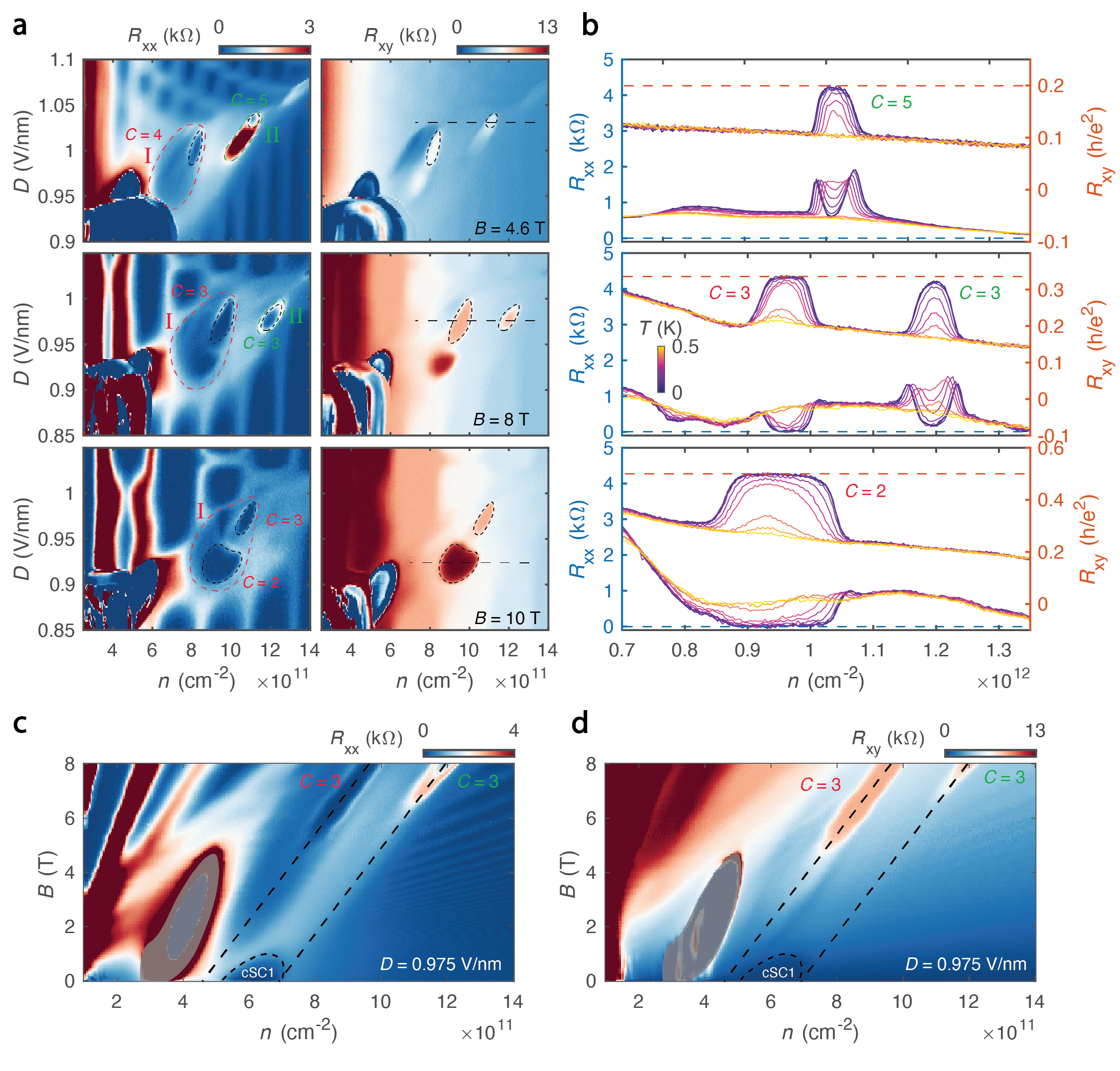} \caption{\textbf{Unconventional RIQH states.} \textbf{a}, $n$--$D$ maps of $R_{xx}$ and $R_{xy}$ at $B = 4.6\,\mathrm{T}$, $B = 8\,\mathrm{T}$ and $B = 10\,\mathrm{T}$. The phase boundaries of regions \RomanNumeralCaps{1} and \RomanNumeralCaps{2} are indicated by red and green dashed lines, respectively. \textbf{b}, Temperature dependence of $R_{xx}$ and $R_{xy}$ at $B = 4.6\,\mathrm{T}$ and $D = 1.034\,\mathrm{V/nm}$, $B = 8\,\mathrm{T}$ and $D = 0.975\,\mathrm{V/nm}$, and $B = 10\,\mathrm{T}$ and $D = 0.925\,\mathrm{V/nm}$. The black dashed lines in the right column of panel \textbf{a} indicate the positions of the temperature measurements at fixed displacement fields. \textbf{c, d}, Landau fan diagram of $R_{xx}$ and $R_{xy}$ at $D = 0.975\,\mathrm{V/nm}$. ($T = 20\,\mathrm{mK}$)}
  \label{fig4}
\end{figure*} 

\Needspace{6\baselineskip}
\section{Unconventional RIQH states}

We have already shown that RIQH states emerge in the lower part of region \RomanNumeralCaps{1} and that they trace back to the WC boundary as indicated in Figs.~\ref{fig2}b with black dashed lines. We now investigate these RIQH states in detail. Our measurements show that the RIQH states originate from two distinct sources and reveal that the system exhibits distinct behaviors under different displacement fields. Fig.~\ref{fig4}a shows the $n$--$D$ maps of $R_{xx}$ and $R_{xy}$ at $B = 4.6\,\mathrm{T}$, $B = 8\,\mathrm{T}$, and $B = 10\,\mathrm{T}$. Region\,\RomanNumeralCaps{1}, which is enclosed by a red dashed line, and region \RomanNumeralCaps{2}, which is enclosed by a green dashed line, shift toward lower displacement field and higher carrier density when the magnetic field is increased. Several RIQH states emerge inside regions \RomanNumeralCaps{1} and \RomanNumeralCaps{2}. Hereafter, we label these RIQH states by their Hall conductance in units of $e^2/h$ as well as based on whether they reside inside of region \RomanNumeralCaps{1} or \RomanNumeralCaps{2}. At $B = 4.6\,\mathrm{T}$, we observe a $C^{\mathrm{\RomanNumeralCaps{1}}}_{4.6}=4$ state in region \RomanNumeralCaps{1} and a $C^{\mathrm{\RomanNumeralCaps{2}}}_{4.6}=5$ state in region \RomanNumeralCaps{2}. The former belongs to the series of RIQH states emerging from the WC phase boundary as shown in Fig.~\ref{fig2}b, whereas the latter appears only within a narrow displacement-field and magnetic-field window in region \RomanNumeralCaps{2} (Fig.~\ref{E5}b). At $B=8\,\mathrm{T}$, two $C^{\mathrm{\RomanNumeralCaps{1}/\mathrm{\RomanNumeralCaps{2}}}}_{8}=3$ states are observed: one in region \RomanNumeralCaps{1} and the other in region \RomanNumeralCaps{2}. Figures~\ref{fig4}c,d show the corresponding Landau fan diagrams at $D=0.975\,\mathrm{V/nm}$. Similar to Fig.~\ref{fig1}b, the $C^{\mathrm{\RomanNumeralCaps{1}}}_{8}=5$ state in region \RomanNumeralCaps{1} emerges from the WC boundary, although only a single robust state is observed here. In contrast, the $C^{\mathrm{\RomanNumeralCaps{2}}}_{8}=5$ state in region \RomanNumeralCaps{2} originates from the boundary of the cSC1 phase rather than the WC phase, as indicated by the right black dashed lines in Figs.~\ref{fig4}c,d. At $B=10\,\mathrm{T}$, region \RomanNumeralCaps{2} is completely suppressed, and no RIQH states are observed there. Instead, two states, $C^{\mathrm{\RomanNumeralCaps{1}}}_{10}=3$ and $C^{\mathrm{\RomanNumeralCaps{1}}}_{10}=2$, appear in region \RomanNumeralCaps{1} at different displacement fields. The $C^{\mathrm{\RomanNumeralCaps{1}}}_{10}=2$ state emerges as the $C^{\mathrm{\RomanNumeralCaps{1}}}_{10}=3$ state disappears, and both states originate from the WC boundary (Fig.~\ref{E5}a). Fig.~\ref{fig4}b shows the temperature dependence of $R_{xx}$ and $R_{xy}$ for these states. The critical temperature of these states is much lower than the conventional quantum Hall edge states, as shown in Fig.~\ref{E6}, and the Hall resistance returns to its normal, non-quantized value at elevated temperatures.

RIQH states have previously been observed in high Landau levels of graphene and GaAs-based systems under strong magnetic fields~\cite{lozovik1975crystallization, andrei1988observation, goldman1990evidence, Eisenstein2002, li2010obseravtion, chen2019competing, tsui2024direct}, where they are generally attributed to the existence of localized charges in bubble phases or Wigner crystal phases~\cite{fogler1996ground, koulakov1996charge, moessner1996exact}. This causes the Hall resistance to reenter quantized values associated with nearby integer filling factors. In our case, however, the Hall resistance does not revert to the nearest integer quantized values. Additionally, as mentioned above, the RIQH states in region \RomanNumeralCaps{1} originate from the boundary of WC phase that already exist at zero magnetic field. These distinctions further highlight the unconventional character of the observed RIQH states. 

Focusing on region \RomanNumeralCaps{1}, these RIQH states most likely can be interpreted as originating from an electron-doped mWC phase. We have already identified a hole-doped mWC phase near the upper boundary of the WC phase (see Fig.~\ref{fig1}). As the carrier density is further increased and reaches region \RomanNumeralCaps{1}, the emergence of an electron-doped mWC (e-mWC) phase may be anticipated~\cite{kim2024dynamical,dong2026crystals,feng2026self}. Under an applied magnetic field, these itinerant carriers in e-mWC then form Landau levels, giving rise to the observed RIQH states. The series of RIQH states shown in Fig.~\ref{fig2}b tracing back to carrier densities associated with the WC phase boundary is consistent with this interpretation. The disagreement between the Hall resistance and the St\v{r}eda formula may then originate from changes in the density of localized carriers. We extract an effective itinerant carrier density from the quantized Hall resistance. We find that, within the RIQH states, the itinerant component corresponds to approximately 50\% of the total carrier density. As discussed above and shown in Fig.~\ref{fig4}c,d, when the displacement field is reduced, the series of RIQH states disappears, leaving only a single robust state with $C^{\mathrm{\RomanNumeralCaps{1}}}=3$ or $C^{\mathrm{\RomanNumeralCaps{1}}}=2$, depending on the displacement field. This behavior can potentially be attributed to a competition between the RIQH states and neighboring phases at low magnetic fields, or these isolated robust $C^{\mathrm{\RomanNumeralCaps{1}}}=3$ and $C^{\mathrm{\RomanNumeralCaps{1}}}=2$ states at low displacement field correspond to particularly stable configurations within the Landau level picture, although the microscopic mechanism underlying the preferential stabilization of different RIQH states at different displacement fields remains unknown. Alternatively, these states may be consistent with a magnetic-field-stabilized topological Wigner crystal scenario, in which the quantized Hall response originates from a nontrivial many-body Chern number associated with the Wigner crystal itself. Such a possibility could emerge from the interplay between strong electronic interactions and nontrivial band topology in RG systems~\cite{tevsanovic1989hall, zhou2024fractional, soejima2024anomalous, dong2024anomalous, dong2024stability,zeng2024sublattice, tan2024parent, kwan2025moire, valenti2025quantum, desrochers2026electronic}.

Finally, we want to comment on the RIQH states that appear in region \RomanNumeralCaps{2}. As mentioned above, the $C^{\mathrm{\RomanNumeralCaps{2}}}_{8}=3$ state can be traced back to the boundary of the cSC1 phase, as shown in Fig.~\ref{fig4}d. The origin of the $C^{\mathrm{\RomanNumeralCaps{2}}}_{4.6}=5$ is harder to extrapolate, since it only appears in a small region of the \textit{n--B--D}-phase space. Fig.~\ref{E7} shows the differential resistance $dV_{xx}/dI$ as a function of DC bias current $I_\mathrm{dc}$ and carrier density $n$ at $D = 1.05\,\mathrm{V/nm}$ that runs through region \RomanNumeralCaps{2}. The strong nonlinear behavior observed in region \RomanNumeralCaps{2} hints a pinning mechanism, suggesting that a charge-ordered state already exist at zero and low magnetic fields. Understanding these RIQH states appearing in the phase diagram requires further investigation.

\Needspace{6\baselineskip}
\section*{Discussion and Conclusion}

In summary, we report four closely intertwined emergent phases in R5G. We first establish a hole-doped mWC phase emerging along the boundary of a WC phase. Second, under an out-of-plane magnetic field, we observe an fSC phase in the upper part of region \RomanNumeralCaps{1} and unconventional RIQH states in the lower part of region \RomanNumeralCaps{1}. Notably, the phase boundary between the mWC and WC phases evolves continuously into the boundary separating the superconducting and RIQH phases, and all four states exhibit comparable critical temperatures. These observations suggest an intimate connection among the emergent phases. At high displacement fields, the RIQH states in the lower part of region \RomanNumeralCaps{1} can be interpreted as emerging from the Landau levels of an electron-doped mWC phase, revealing the potential realization of both electron- and hole-doped polarities of the mWC phase in R5G. Future experiments that probe the local charge texture via scanning tunneling microscopy will be highly valuable to gain a deeper understanding of the mWC phase. In parallel to experimental investigations, further theoretical developments will aid in understanding the unique properties of these self-doped itinerant carriers. 

The coexistence of the fSC and RIQH states raises the intriguing possibility that the superconducting state may also inherit an underlying crystalline background. Specifically, Cooper pairing might occur between the self-doped itinerant carriers of the mWC rather than involving the entire electron fluid without localized carriers. The exact nature of this fSC phase, its pairing mechanism, and its relationship with the surrounding electronic phases incentivizes further study. Regarding the low-displacement-field $C^{\mathrm{\RomanNumeralCaps{1}}}=2$ and $C^{\mathrm{\RomanNumeralCaps{1}}}=3$ states, tracking their excitation gaps versus magnetic field via quantum capacitance measurements can help to clarify their nature. Within a conventional Landau-level framework, and assuming that the self-doped carriers behave normally under a magnetic field, the energy gap scales linearly with the field. Significant nonlinear deviations may thus signal a topologically ordered state. 

Finally, we emphasize that the R5G system remains far from fully understood within the regime of high displacement fields and low carrier densities. The phase diagram is rich, and its constituent phases appear to be continuously connected. What our results suggest is that the Wigner crystal is not merely an insulating state in the R5G phase diagram, but may instead serve as a crystalline background from which a remarkable variety of correlated phases emerges. We hope this perspective will motivate further experimental and theoretical efforts to understand the nature and origin of these intertwined phases.

\newpage
\clearpage

\bibliography{R5G}

\section*{METHODS}

\subsection{Sample fabrication and Measurements}

The material stacks are fabricated using a dry-transfer technique based on a two-step procedure. First, we prepare the bottom part of the device, consisting of the bottom hBN and the bottom gate. To ensure a clean and exposed surface of the bottom hBN, we employ a recently established “dragging” method. Specifically, a hBN/hBN/bottom-gate stack is first assembled and then released onto SiO$_2$-coated Si-wafer chips with alignment markers. The top hBN is subsequently dragged away, leaving the top surface of the bottom hBN clean and free of contamination. In the second step, the top stack, which comprises the top gate, top hBN, and rhombohedral graphene, is picked up and subsequently transferred onto the assembled bottom part. 5nm/15nm/70nm Cr/Pd/Au is evaporated to electrically connect the graphene via one-dimensional contacts.

Electronic transport measurements were performed in four different cryogenic systems: an Oxford He-3 refrigerator with a base temperature of $T = 240\,\mathrm{mK}$, a Bluefors dilution refrigerator with a base temperature of $T = 11\,\mathrm{mK}$, a Leiden MCK dilution refrigerator with a base temperature of $T = 50\,\mathrm{mK}$, and a Cryoconcept Hexadry200C system with a base temperature of $T = 20\,\mathrm{mK}$. Resistance measurements were carried out using a standard lock-in technique with an AC excitation current of $I = 1\text{--}5\,\mathrm{nA}$ at a frequency of $f = 17.7777\,\mathrm{Hz}$. Top and bottom gate voltages were applied using Yokogawa GS200 source meters to tune the carrier density and displacement field. The voltage signals were amplified by a factor of 100 before acquisition. The carrier density and displacement field were determined according to $n=(C_\text{TG}V_\text{TG}+C_\text{BG}V_\text{BG})/e$ and $D/\varepsilon_0=(C_\text{TG}V_\text{TG}-C_\text{BG}V_\text{BG})/2\varepsilon_0$, respectively, where $C_\text{TG/BG}$ and $V_\text{TG/BG}$ are the top and bottom gate capacitances per unit area and voltages, respectively. The gate capacitances depend on the hBN layer thicknesses and were determined via Hall resistance measurements at different gate-defined carrier densities.

\begin{acknowledgments}
We acknowledge helpful discussions with S Kivelson, S. A. Parameswaran, Y. H. Kwan, C Lewandowski, and T. Neupert. Z.Z. acknowledges the help from B. Ghosh and A. K. Paul with the measurements in WIS. Z.Z. and K.K. acknowledge funding from SNSF. S.J. acknowledges support from Florida State  University through the Quantum Postdoctoral Fellowship and the National High Magnetic Field Laboratory. The National High Magnetic Field Laboratory is supported by the National Science Foundation through NSF/DMR-2128556 and the State of Florida. K.W. and T.T. acknowledge support from the JSPS KAKENHI (Grant Numbers 20H00354 and 23H02052) and World Premier International Research Center Initiative (WPI), MEXT, Japan. M.H. acknowledges support of the Israel Science Foundation, Grant No. 1510/22. R.A. and P.R. acknowledge support from ERC starting grant COHEGRAPH 679531, ANR EQUBITFLY, Horizon-EIC-2022-Pathfinderopen project FLATS, and Horizon-EIC-2024-Pathfinderopen project ELEQUANT. M.B. acknowledges the support of the SNSF Eccellenza grant No. PCEGP2\_194528.

\end{acknowledgments}

\textbf{\begin{center}Author Contributions\end{center}}

M.B. supervised the project. Z.Z., K.K., and Y.L fabricated the devices. Z.Z. and K.K. performed the measurements in EPFL; Z.Z. performed the measurements in WIS; Z.Z. and R.A. performed the measurements in CEA. S.J. commented on and provided suggestions for the manuscript. T.T. and K.W. provided the hBN crystal. M.H. and P.R. collaborated on discussions and analysis and provided the wet ($T = 50\,\mathrm{mK}$, WIS) and dry dilution ($T = 20\,\mathrm{mK}$, CEA) fridges for the measurements. Z.Z. analyzed the data and wrote the manuscript with input from all authors.

\textbf{\begin{center}Competing interests\end{center}}
The authors declare no competing interests.

\textbf{\begin{center}Data availability\end{center}}
The data supporting the findings of this study are available from the corresponding author upon reasonable request.

\newpage
\clearpage
\onecolumngrid

\section*{Extended figures}

\begin{figure*}[!htbp]
  \centering
  \renewcommand{\thefigure}{E1}
  \includegraphics[width= 0.75\textwidth]{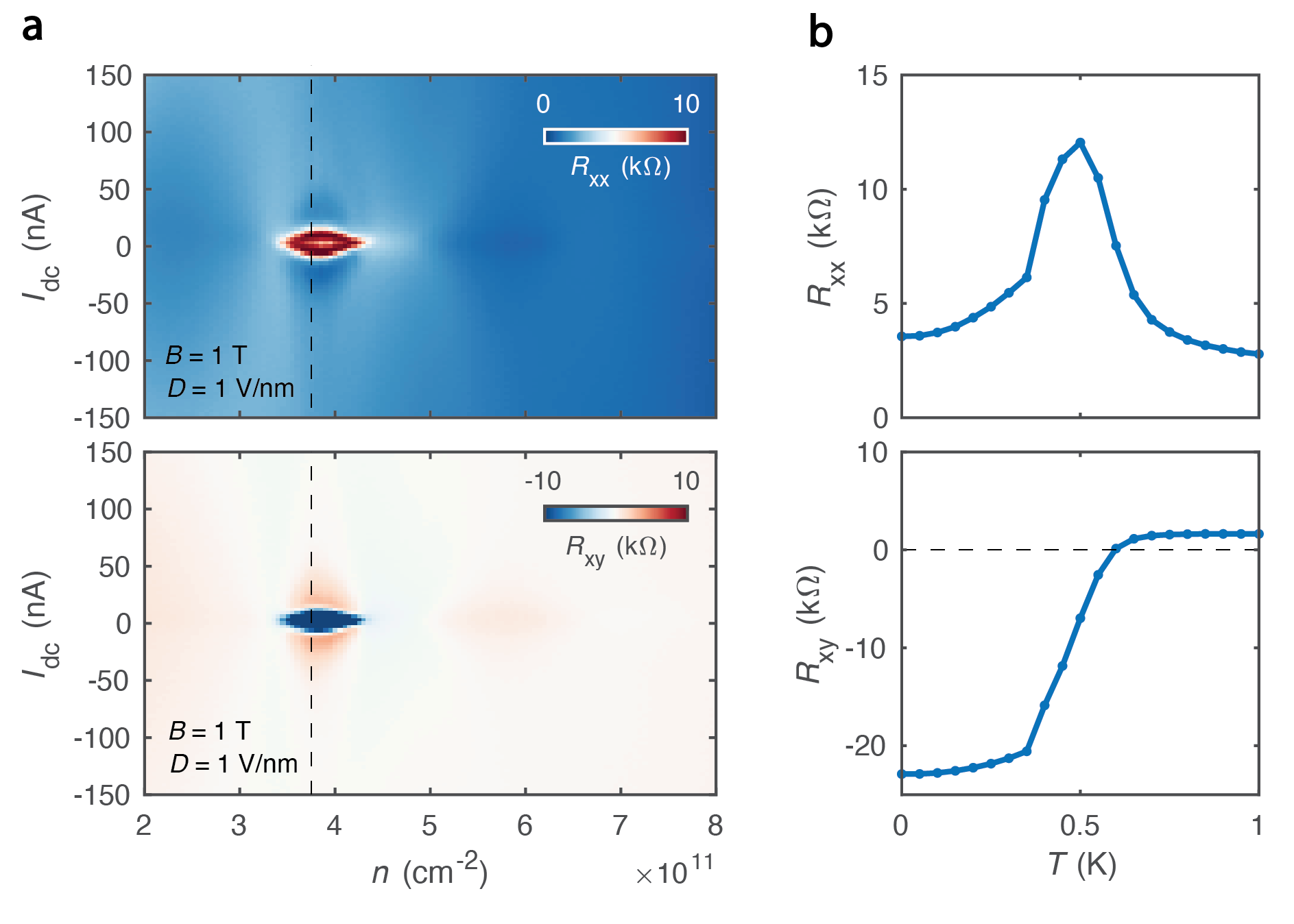} \caption{\textbf{Characterization of the mWC state.} \textbf{a,} $dV_{xx}/dI$ and $dV_{xy}/dI$ versus DC current $I_{dc}$ and carrier density $n$ at a fixed displacement field of $D = 1\,\mathrm{V/nm}$ and $B = 1\,\mathrm{T}$. Both $V_{xx}$ and $V_{xy}$ show strong nonlinear behavior as a function of the DC bias current, $I_{\mathrm{dc}}$. With increasing $I_{\mathrm{dc}}$, the Hall resistance changes sign at high bias. \textbf{b,} Temperature dependence of $R_{xx}$ and $R_{xy}$ of the mWC phase. A pronounced peak in $R_{xx}$ and a concomitant sign change in $R_{xy}$ are observed with increasing temperature.}
  \label{E1}
\end{figure*}

\begin{figure*}
  \centering
  \renewcommand{\thefigure}{E2}
  \includegraphics[width= 0.85\textwidth]{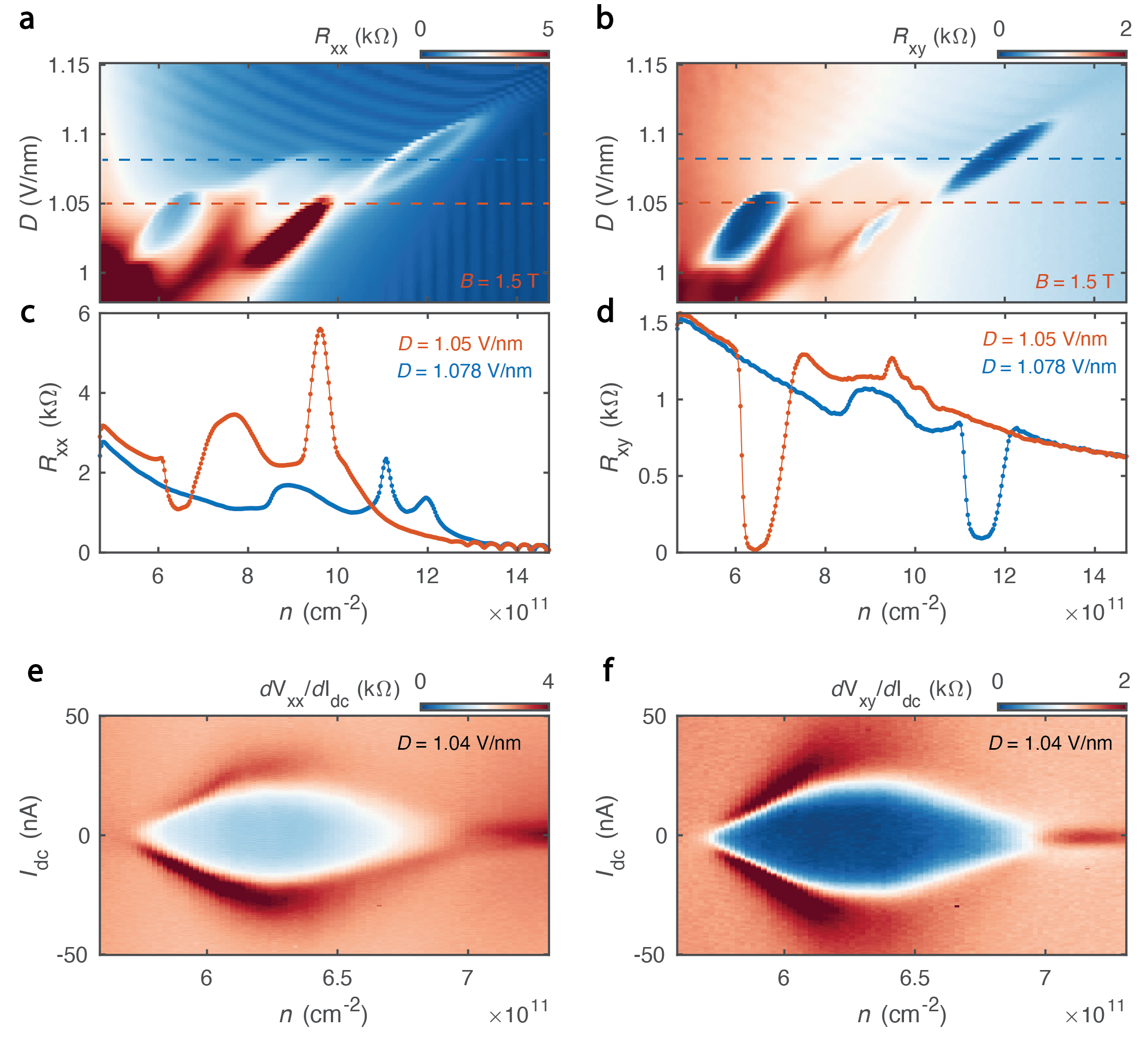} \caption{\textbf{Magnetic-field-stabilized superconductivity in R5G\_2.} \textbf{a, b,} $n$--$D$ maps of $R_{xx}$ and $R_{xy}$ measured at $B = 1.5\,\mathrm{T}$. \textbf{c, d,} Linecuts of $R_{xx}$ and $R_{xy}$ at $D = 1.05\,\mathrm{V/nm}$ (orange) and $D = 1.078\,\mathrm{V/nm}$ (blue). We can see a drop in $R_{xx}$ and $R_{xy}$ in regions \RomanNumeralCaps{1} and \RomanNumeralCaps{3}. In region \RomanNumeralCaps{1}, $R_{xy}$ drops to zero while $R_{xx}$ retains a finite residual value. Here, $R_{xx}$ is higher than what we have shown in the main text. This is because the distance between the two voltage probes used for $R_{xx}$ is larger than that in R5G\_1. \textbf{e, f,} $dV_{xx}/dI$ and $dV_{xy}/dI$ versus DC bias current $I_{dc}$ and carrier density $n$ at a fixed displacement field of $D = 1.04\,\mathrm{V/nm}$ and $B = 1.5\,\mathrm{T}$. Nonlinear behavior exists in region \RomanNumeralCaps{1} (The data were acquired at $T = 50\,\mathrm{mK}$).}
  \label{E2}
\end{figure*} 

\begin{figure*}
  \centering
  \renewcommand{\thefigure}{E3}
  \includegraphics[width= 0.95\textwidth]{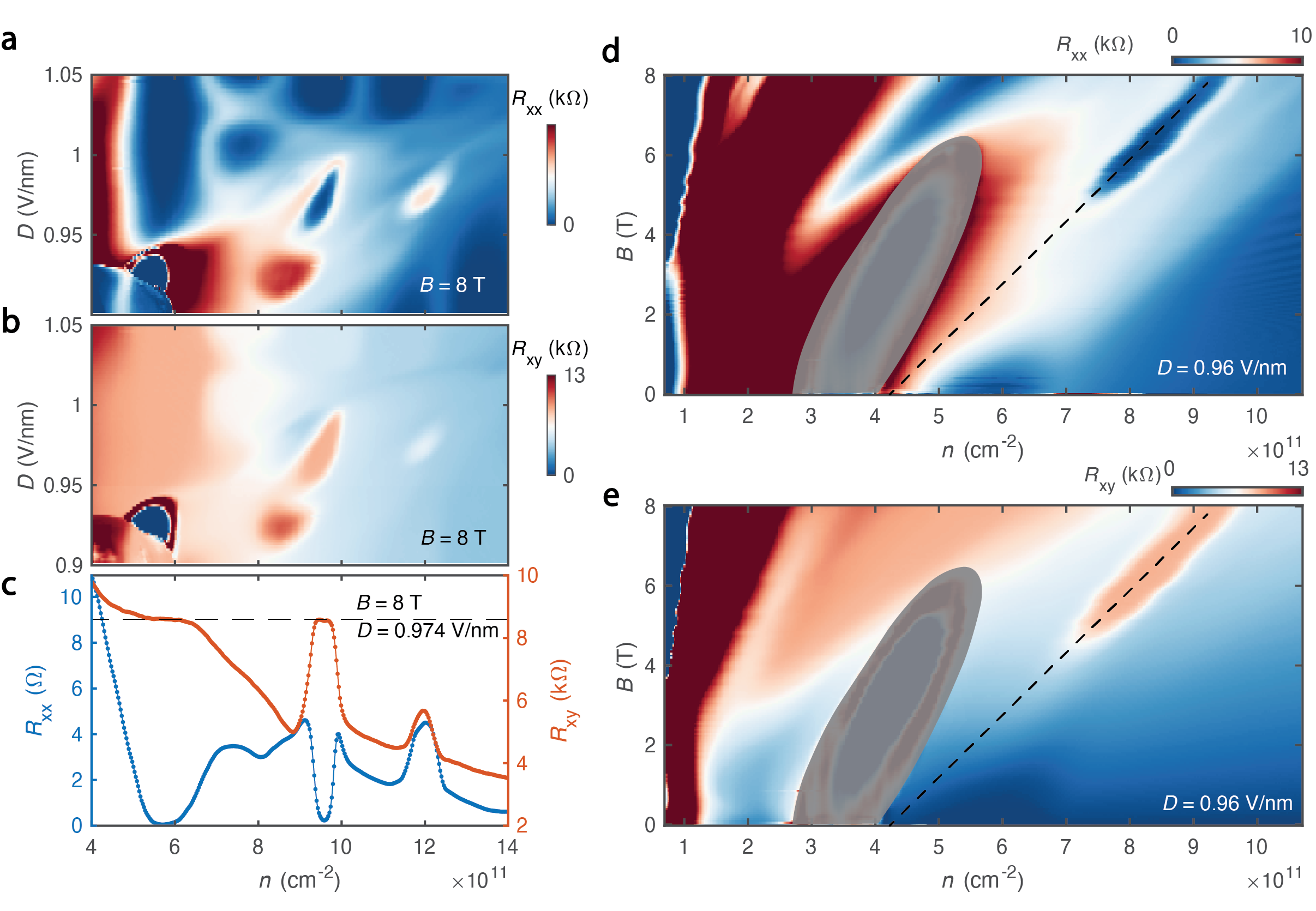} \caption{\textbf{$C = 3$ state in R5G\_2.} \textbf{a, b,} $n$--$D$ maps of $R_{xx}$ and $R_{xy}$ measured at $B = 8\,\mathrm{T}$. A “bubble” region is visible with vanishing $R_{xx}$ and quantized $R_{xy}$. \textbf{c,} Line cuts of $R_{xx}$ (blue) and $R_{xy}$ (orange) extracted from \textbf{a} and \textbf{b} at $D = 0.974\,\mathrm{V/nm}$. The black line serves as a guide to the eye for the quantized Hall resistance $h/3e^2$. \textbf{d, e,} Landau fan diagrams of $R_{xx}$ and $R_{xy}$ at $D = 0.96\,\mathrm{V/nm}$. The black dashed lines trace the state exhibiting a quantized Hall resistance of $h/3e^2$. Similar to what we have shown in the main text, this $C=3$ state traces back to the boundary of the WC phase at zero field. (The data were acquired at $T = 50\,\mathrm{mK}$).}
  \label{E3}
\end{figure*} 

\begin{figure*}
  \centering
  \renewcommand{\thefigure}{E4}
  \includegraphics[width= 0.85\textwidth]{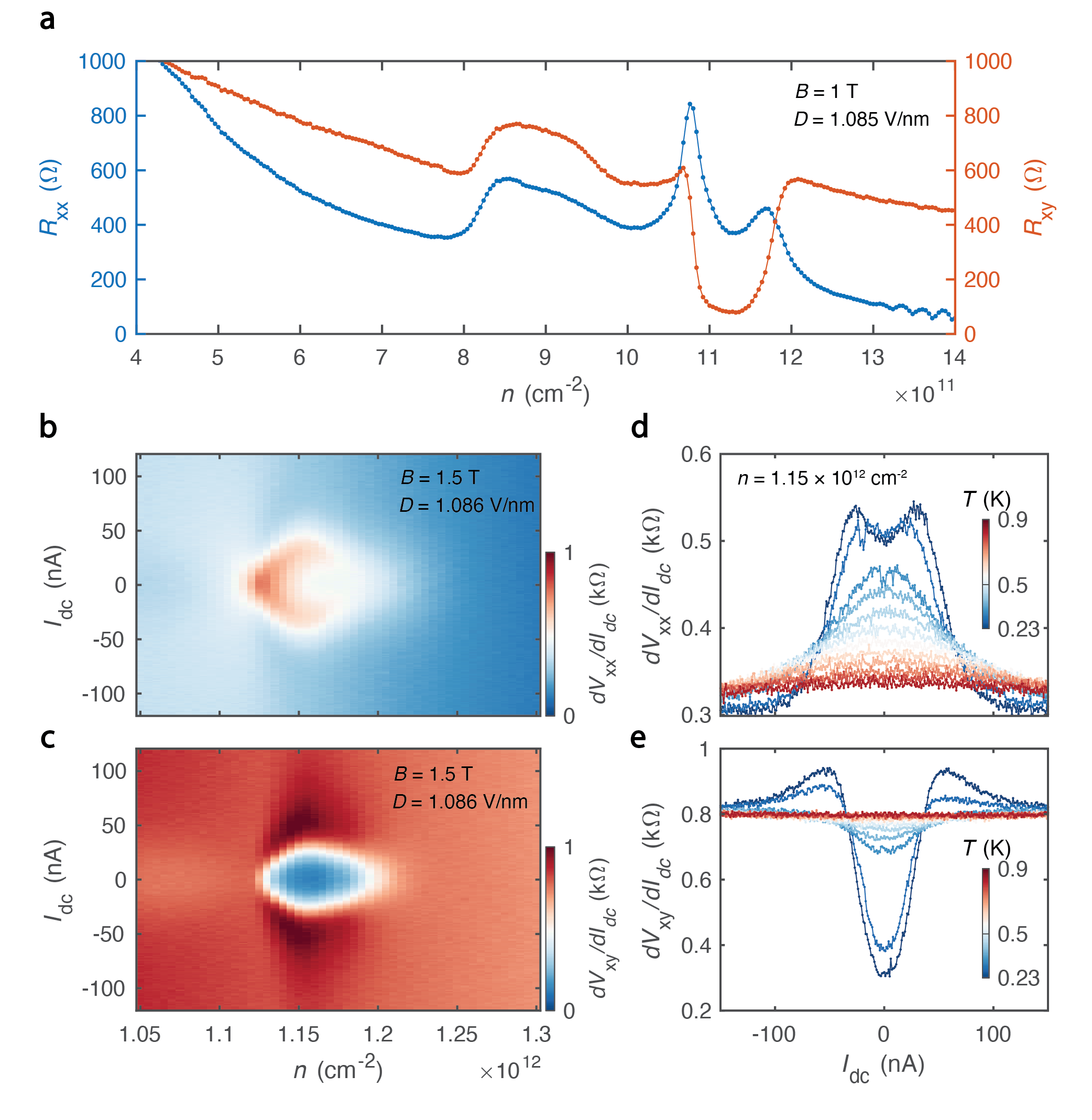} \caption{\textbf{Unconventional behavior in region \RomanNumeralCaps{3}.} \textbf{a,} Linecuts of $R_{xx}$ and $R_{xy}$ at $D = 1.085\,\mathrm{V/nm}$ and $B = 1\,\mathrm{T}$. $R_{xy}$ shows a sudden drop from the normal Hall resistance around $n = 1.12 \times 10^{12}\,\mathrm{cm^{-2}}$ and $R_{xx}$ also shows a small dip. This state may also be an out-of-plane magnetic-field-stabilized superconducting phase. \textbf{b, c,} $dV_{xx}/dI$ and $dV_{xy}/dI$ as a function of DC current $I_{dc}$ and carrier density $n$ at a fixed displacement field of $D = 1.086\,\mathrm{V/nm}$ and $B = 1.5\,\mathrm{T}$. It is clear that nonlinear behavior emerges in the region where both $R_{xx}$ and $R_{xy}$ exhibit a drop. \textbf{d, e,} Temperature dependence of $dV_{xx}/dI$ and $dV_{xy}/dI$ at $D = 1.086\,\mathrm{V/nm}$ and $B = 1.5\,\mathrm{T}$ and $n = 1.15 \times 10^{12}\,\mathrm{cm^{-2}}$. The nonlinear behavior can be suppressed by increasing the temperature, and the system eventually transitions to an ohmic linear region (The data were acquired at $T = 240\,\mathrm{mK}$).}
  \label{E4}
\end{figure*}

\begin{figure*}
  \centering
  \renewcommand{\thefigure}{E5}
  \includegraphics[width= 0.95\textwidth]{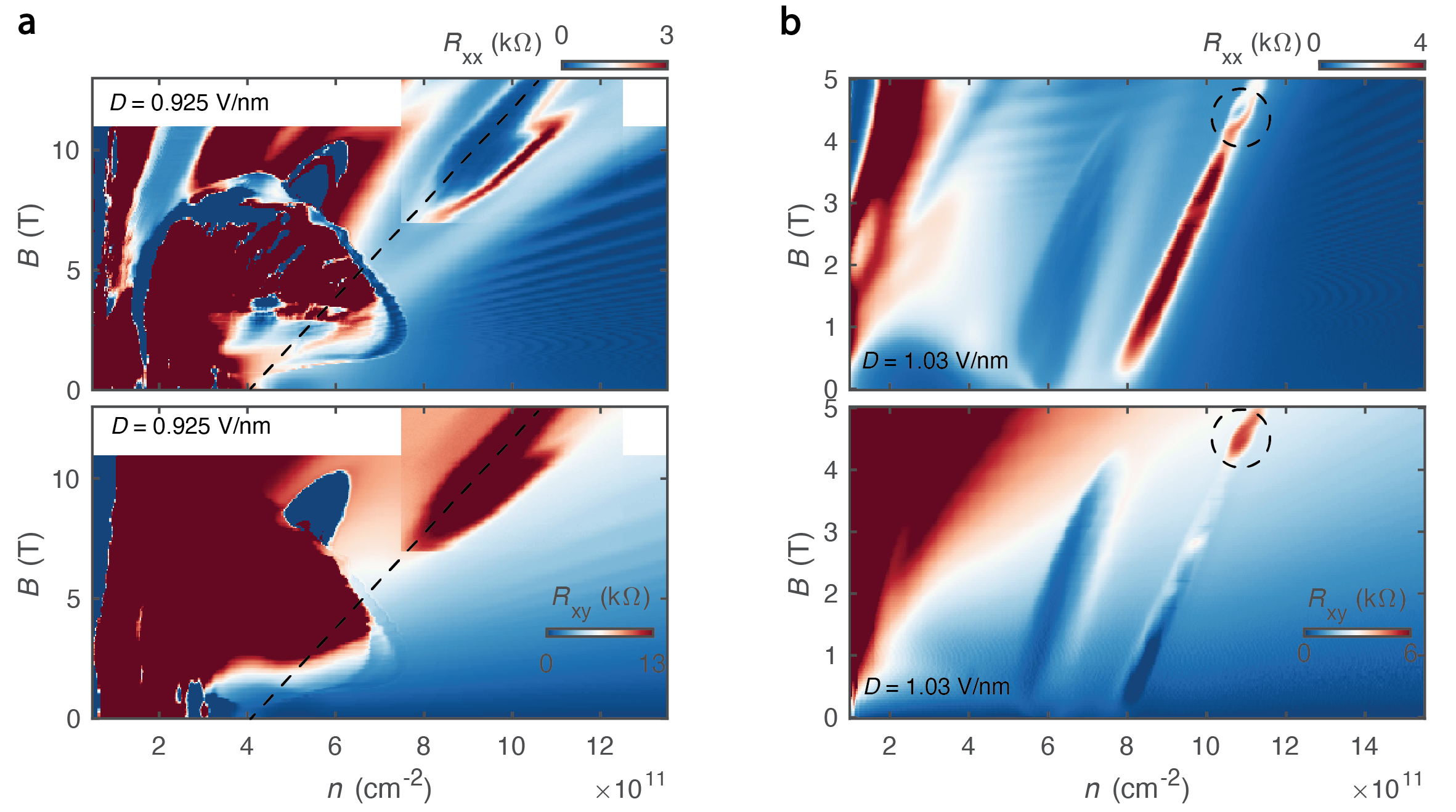} \caption{\textbf{Magnetic field dependence of the $C^{\mathrm{\RomanNumeralCaps{1}}}=2$ and $C^{\mathrm{\RomanNumeralCaps{2}}}=5$ states.} \textbf{a,} Landau fan diagrams of $R_{xx}$ and $R_{xy}$ at $D = 0.925\,\mathrm{V/nm}$. The black dashed lines marks the $C^{\mathrm{\RomanNumeralCaps{1}}}=2$ state. The $C^{\mathrm{\RomanNumeralCaps{1}}}=2$ state also stems from the boundary of WC phase at zero magnetic field. (Here the main fan diagram data is taken at $T = 240\,\mathrm{mK}$ and the zoomed-in fan diagram is taken at $T = 140\,\mathrm{mK}$). \textbf{b,} Landau fan diagrams of $R_{xx}$ and $R_{xy}$ at $D = 1.03\,\mathrm{V/nm}$ and at $T = 240\,\mathrm{mK}$. The black dashed circle marks the region where $C^{\mathrm{\RomanNumeralCaps{2}}}=5$ state appears.}
  \label{E5}
\end{figure*}

\begin{figure*}
  \centering
  \renewcommand{\thefigure}{E6}
  \includegraphics[width= 0.85\textwidth]{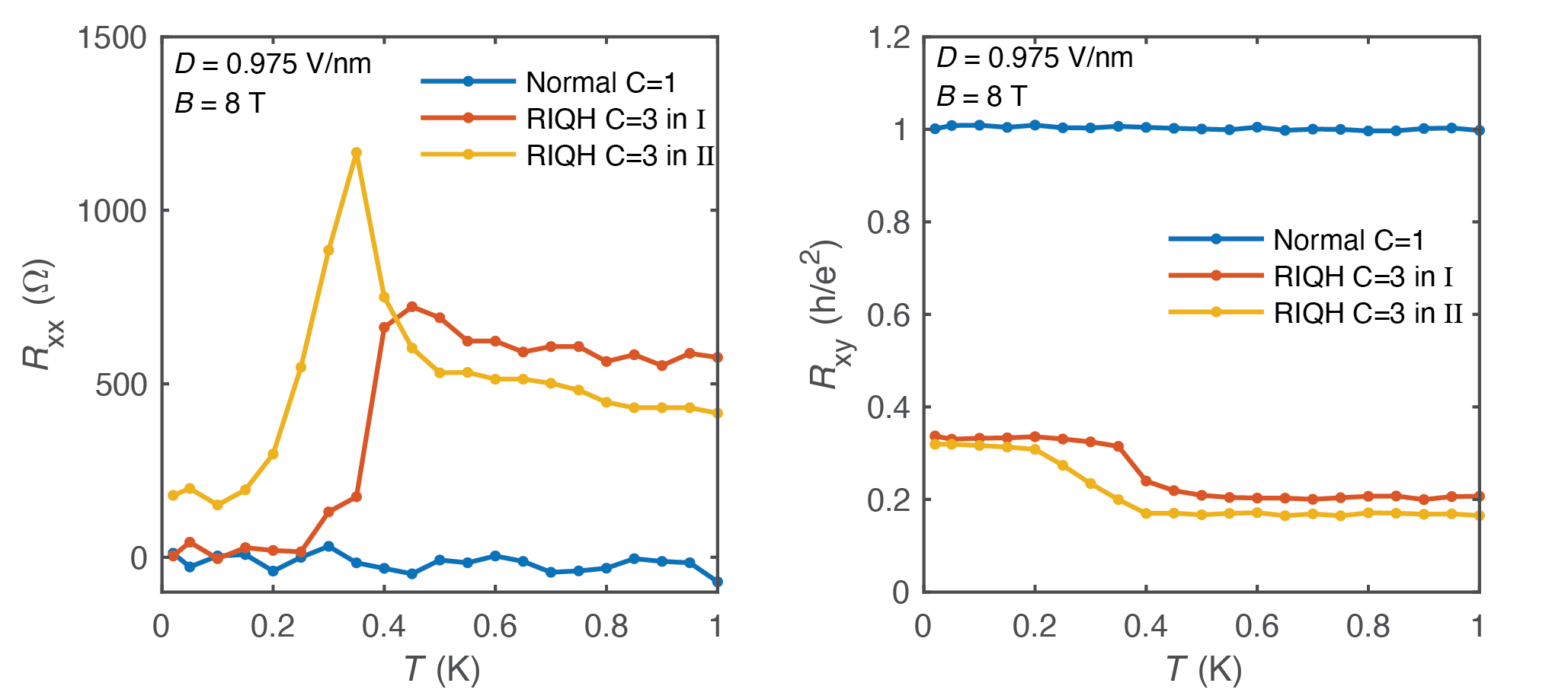} \caption{\textbf{Temperature dependence of normal Landau level and RIQH states.} $R_{xx}$ and $R_{xy}$ as a function of temperature of three different states at $D = 0.975\,\mathrm{V/nm}$, $B = 8\,\mathrm{T}$. Blue: normal Landau level with $C=1$; orange: RIQH state with $C^{\mathrm{\RomanNumeralCaps{1}}}=3$ in region \RomanNumeralCaps{1}; yellow: RIQH state with $C^{\mathrm{\RomanNumeralCaps{2}}}=3$ in region \RomanNumeralCaps{2}.}
  \label{E6}
\end{figure*}

\begin{figure*}
  \centering
  \renewcommand{\thefigure}{E7}
  \includegraphics[width= 0.95\textwidth]{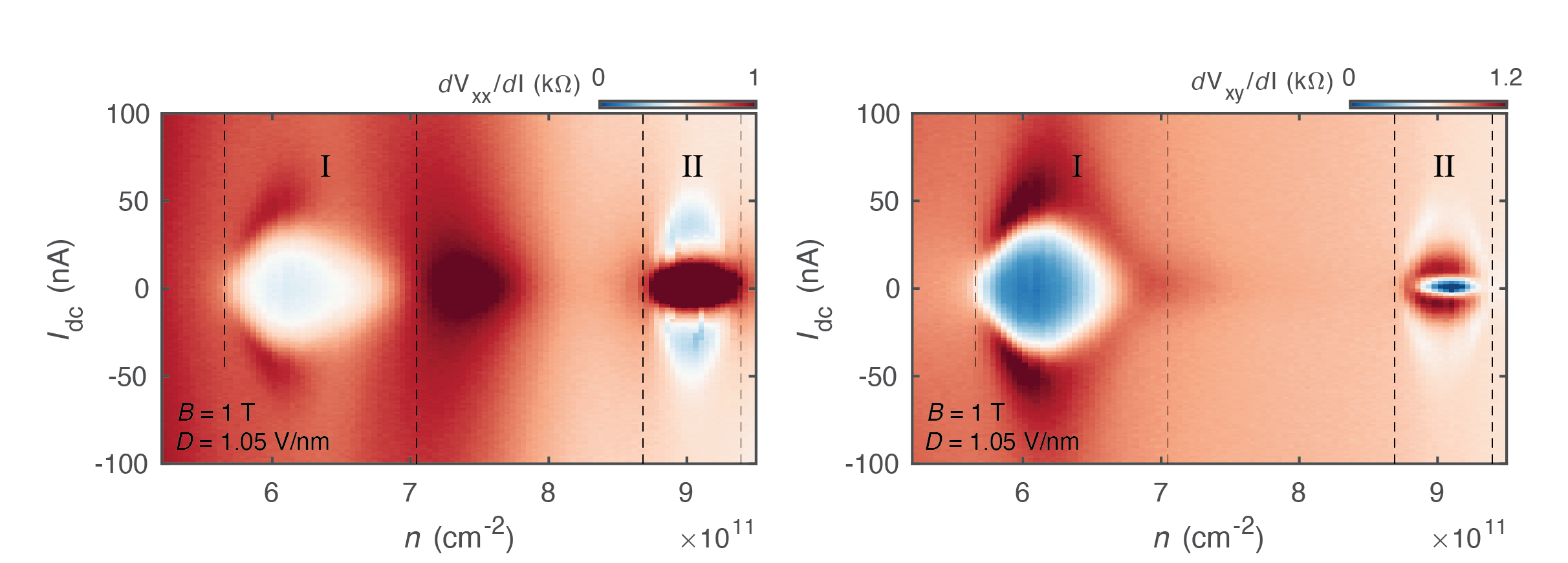} \caption{\textbf{Nonlinear $V$--$I$ behavior in region \RomanNumeralCaps{2}.} Differential resistances $dV_{xx}/dI$ and $dV_{xy}/dI$ as functions of DC bias current $I_{dc}$ and carrier density $n$, measured at a fixed displacement field of $D = 1.05\,\mathrm{V/nm}$ and a magnetic field of $B = 1\,\mathrm{T}$. In addition to region \RomanNumeralCaps{1}, region \RomanNumeralCaps{2} also exhibits pronounced nonlinear $V$--$I$ behavior, suggesting the presence of a charge-ordered state. (The data were acquired at $T = 240\,\mathrm{mK}$)}
  \label{E7}
\end{figure*}

\end{document}